\documentclass[preprint,12pt]{elsarticle}

\usepackage{amssymb}
\usepackage{amsmath}
\usepackage{graphicx}
\usepackage{booktabs}
\usepackage{float}
\usepackage[ruled,vlined]{algorithm2e}
\usepackage{hyperref}
\usepackage{xcolor}

\newcounter{bla}

\journal{Computer Physics Communications}

\begin{document}

\begin{frontmatter}

\title{MLM: Multi-Layer Moir\'{e} --- A Python Package for
       Generating Commensurate Supercells of Twisted Multilayer
       Two-Dimensional Materials}

\author[a]{Anikeya Aditya\corref{cor1}}
\author[a,b]{Sampad Mohanty}

\cortext[cor1]{Corresponding author.\\
  \textit{E-mail address:} anikeya9@gmail.com}

\begin{abstract}
Moir\'{e} superlattices formed by stacking atomically thin
two-dimensional materials with a relative twist angle have emerged as
a versatile platform for engineering quantum electronic, optical, and
ferroic properties. Computational modelling of such systems with
periodic boundary conditions requires the identification of
commensurate supercells in which the moir\'{e} periodicity is
reproduced exactly, or within a prescribed tolerance. While several
codes exist for bilayer systems, extension to three or more layers
with independently chosen twist angles remains a significant
challenge. Here we present \textbf{MLM} (Multi-Layer Moir\'{e}),
an open-source Python package that constructs periodic,
PBC-compatible moir\'{e} supercells for an arbitrary number of
twisted layers with any Bravais lattice type. The package employs a
solve-and-round algorithm that reduces the coincidence-site search to
an $\mathcal{O}(N^2)$ linear-algebra problem per twist angle,
compared to the $\mathcal{O}(N^4)$ brute-force enumeration required
by conventional approaches. We demonstrate the package on bilayer
graphene, bilayer and trilayer MoS$_2$, bilayer SrTiO$_3$, and a
PbTiO$_3$/BaTiO$_3$ oxide heterostructure, producing
simulation-ready structure files for both VASP and LAMMPS. The
fractional-coordinate atom-selection algorithm scales to supercells
containing millions of atoms and is robust across all twist angles
including very small angles below $1^\circ$.

\bigskip
\noindent\textbf{PROGRAM SUMMARY}

\begin{small}
\noindent
{\em Program title:} MLM (Multi-Layer Moir\'{e})\\
{\em CPC Library link to program files:} (to be assigned by Technical Editor)\\
{\em Developer's repository:}
  \url{https://github.com/anikeya9/Multi-Layer-Moire-MLM-/tree/cleaned-up}\\
{\em Licensing provisions:} MIT\\
{\em Programming language:} Python 3 ($\geq$3.9)\\
{\em Supplementary material:} None\\
{\em Nature of problem (approx.\ 100 words):}\\
Constructing periodic boundary condition (PBC)-compatible supercells
for twisted multilayer two-dimensional materials is required for
density functional theory (DFT) and molecular dynamics (MD)
simulations. For an arbitrary twist angle, the moir\'{e} pattern is
generally incommensurate and cannot be simulated directly under
periodic boundary conditions. The problem reduces to finding integer
linear combinations of two (or more) rotated lattice bases that
coincide within a prescribed tolerance — a high-dimensional integer
lattice matching problem. For multilayer stacks, each additional layer
introduces an independent rotational constraint, and the existing
bilayer-focused tools do not directly extend to three or more
layers.\\

{\em Solution method (approx.\ 100 words):}\\
For each candidate twist angle $\theta$ and each integer pair
$(n_1,n_2)$ indexing a bottom-layer lattice vector
$\mathbf{v}=A\,(n_1,n_2)^\top$, MLM solves the $2\times2$ linear
system $B(\theta)\,\mathbf{m}=\mathbf{v}$ to obtain the
corresponding top-layer fractional indices $\mathbf{m}\in\mathbb{R}^2$.
A match is accepted when both components of $\mathbf{m}$ lie within a
user-specified tolerance $\delta$ of the nearest integers. This
solve-and-round strategy requires only $\mathcal{O}(N^2)$ linear
solves per angle, where $N$ is the search range, compared to
$\mathcal{O}(N^4)$ for brute-force enumeration. Multilayer extension
is sequential: supercell vectors found at each stage serve as the
reference lattice for the next layer.\\

{\em Additional comments:}\\
The package supports homobilayers, heterobilayers (different lattice
constants), trilayers, and arbitrary multilayer stacks. Atomic
structure generation uses a fractional-coordinate folding algorithm
that scales to millions of atoms and correctly handles all twist
angles and supercell orientations without polygon-clipping tolerances.
Commensurate structure files are written in VASP POSCAR and LAMMPS
atomic-style data formats via the Atomic Simulation Environment
(ASE).\\

\end{small}

\end{abstract}

\begin{keyword}
  moiré superlattice \sep
  twisted trilayer \sep
  commensurate supercell \sep
  coincidence lattice \sep
  van der Waals heterostructure \sep
  periodic boundary conditions \sep
  Python
\end{keyword}

\end{frontmatter}

\section{Introduction}
Moiré superlattices formed by twisting or lattice-mismatching atomically thin crystalline layers have emerged as a powerful platform for engineering electronic, optical, and ferroic properties in condensed matter systems. By introducing a relative rotation between layers, a long-wavelength periodic potential—known as the moiré pattern—is generated, leading to substantial reconstruction of electronic band structures and the emergence of collective quantum phenomena. Canonical examples include correlated insulating states and unconventional superconductivity in magic-angle twisted bilayer graphene, as well as moiré excitons and charge localization in transition metal dichalcogenide (TMDC) heterostructures \cite{frontiers_moire_2022,Feuerbacher.2021.ACSAFaA,Naik.2019.TJoPCC} .

Beyond bilayers, recent experimental and theoretical efforts have increasingly focused on \textbf{multilayer moiré systems}, such as twisted trilayer graphene, double bilayer graphene, and more complex heterostructures, where additional rotational degrees of freedom introduce hierarchical moiré patterns and new emergent behavior \cite{frontiers_moire_2022}. In these systems, the interplay between multiple moiré length scales, stacking sequences, and interlayer coupling can stabilize flat electronic bands, topological states, and complex collective phases that are absent in simpler bilayer configurations \cite{frontiers_moire_2022,SanchezSantolino2024}. This growing interest reflects a broader consensus that multilayer moiré platforms provide a qualitatively richer design space for twistronics and synthetic quantum matter \cite{frontiers_moire_2022,Du.2023.S}.

From a computational perspective, predictive modeling of moiré materials relies critically on the ability to construct \textbf{periodic, commensurate supercells} compatible with atomistic simulation techniques such as density functional theory (DFT) and molecular dynamics. For arbitrary twist angles, moiré patterns are generally incommensurate, rendering direct simulation intractable under periodic boundary conditions. As a result, theoretical and computational studies typically focus on special commensurate configurations, where lattice coincidence enables the definition of a finite simulation cell capturing the moiré periodicity.

Several computational tools have been developed to address this challenge for \textbf{bilayer systems}. Notably, the TWISTER package implements coincidence lattice theory to construct commensurate moiré supercells for a wide range of two-dimensional homobilayers and heterobilayers, with optional strain accommodation and structural relaxation workflows . Such approaches have proven highly successful for bilayer graphene, TMDCs, and related van der Waals materials. However, their direct extension to \textbf{three or more layers} is nontrivial. In multilayer stacks, each interface introduces an independent rotational constraint, and enforcing global periodicity across all layers rapidly becomes a high-dimensional integer-lattice problem. Existing bilayer-focused frameworks do not generally guarantee periodic boundary condition (PBC) compatibility across multiple twisted layers, limiting their applicability to emerging multilayer moiré systems.

At the same time, recent studies demonstrate that even subtle changes in stacking configuration and rotational alignment can qualitatively alter the emergent moiré landscape, including the formation of stacking-dependent polarization textures, ferroelectric domains, and rotationally driven symmetry breaking. These findings underscore the need for general-purpose computational tools capable of systematically generating commensurate moiré supercells for multilayer systems with arbitrary lattice types and twist angles.

In this work, we present \textbf{MLM}, a computational framework designed to construct periodic, PBC-consistent moiré supercells for multilayer two-dimensional materials. The method generalizes coincidence-based approaches by explicitly enforcing lattice compatibility across multiple rotated layers using a solve-and-round strategy combined with residual-based validation. The resulting supercells are suitable for direct use in atomistic simulations, enabling systematic exploration of multilayer moiré phenomena beyond the bilayer limit. By decoupling supercell construction from material-specific assumptions, the approach is applicable to a broad class of van der Waals and non–van der Waals systems, including complex oxides and ferroelectric heterostructures .

\section{Nature of the Problem}

The construction of periodic moiré supercells for atomistic simulations
requires identifying lattice vectors that simultaneously satisfy
coincidence conditions between two or more rotated crystalline layers.
For an arbitrary twist angle, two-dimensional lattices are generally
incommensurate, and their moiré pattern has no finite periodicity or the moire lattice vectors are of intractable size. Work by Feurerbacher shows how a small change in twist angles can lead to change of moire lattice vector length by orders of magnitude \cite{Feuerbacher.2021.ACSAFaA}. Practical
simulations therefore focus on special commensurate configurations at which a
finite supercell reproduces the moiré periodicity exactly or within a
prescribed tolerance
\cite{Feuerbacher.2021.ACSAFaA,frontiers_moire_2022,Naik.2022.CPC}.

For bilayer systems with identical hexagonal lattices, analytic expressions or
specialized coincidence-lattice formulations can often determine commensurate
twist angles efficiently \cite{Feuerbacher.2021.ACSAFaA,Naik.2022.CPC}.
However, when moving beyond the highly symmetric homobilayer case — for
instance to heterobilayers with different lattice constants or to systems with
square or rectangular symmetry — the closed-form angle expressions no longer
apply. Extension to three or more layers with independently chosen twist angles
between successive interfaces is substantially more complex still. Each
additional interface introduces an independent rotational constraint, and
enforcing global periodicity requires finding a single superlattice vector pair
that is simultaneously commensurate with all rotated lattices. This reduces to
a higher-dimensional integer lattice matching problem that has no general
closed-form solution.

Existing bilayer-oriented frameworks address this challenge for
two-layer systems using brute-force enumeration of candidate integer index
combinations \cite{Naik.2022.CPC} or symmetry-specific coincidence lattice
approaches \cite{Feuerbacher.2021.ACSAFaA}. These methods scale as
$\mathcal{O}(N^4)$ in the search range $N$, which limits the practically
accessible angle resolution and search range. Their direct extension to
multilayer stacks is also non-trivial: each additional layer multiplies the
search space by an additional factor of $N^2$, making brute-force enumeration
prohibitively expensive for three or more layers.

An additional challenge arises from the asymmetry inherent in the matching
problem. Because the commensurability condition is solved relative to the
bottom layer, the resulting supercell reproduces the bottom-layer periodicity
exactly while the top layer matches it only approximately, with a residual
vector mismatch $\delta_{\mathrm{vec}}$ (denoted \texttt{delvec} in the code)
that depends on both the twist angle and the chosen tolerance. This residual
must be accounted for in the subsequent atomic structure generation step to
avoid incorrect atom placement or double-counting at cell boundaries.

The core problem addressed by MLM is therefore: given a set of two-dimensional
primitive lattice bases $\{A_i\}$ and prescribed relative twist angles
$\{\theta_i\}$ applied sequentially between layers, determine supercell lattice
vectors that enforce global PBC compatibility across all layers, and produce
corresponding atomic structure files directly suitable for atomistic
simulations. The solution must be efficient enough to scan dense angle grids
over the full range from $0^\circ$ to $90^\circ$ on a single workstation, and
robust enough to handle arbitrary lattice types including perovskite oxides with
tetragonal unit cells.

\section{Solution Method Overview}

MLM constructs commensurate moiré supercells using a solve-and-round
coincidence-search strategy that is independent of lattice symmetry and
directly applicable to arbitrary two-dimensional Bravais lattices, including
hexagonal, square, rectangular, and oblique systems.

\subsection{Core idea: solve for top-layer indices}

The key observation is that a brute-force search need not enumerate all four
indices $(n_1, n_2, m_1, m_2)$ simultaneously. Instead, for each candidate
bottom-layer vector $\mathbf{v} = A\,(n_1,n_2)^\top$, the corresponding
top-layer fractional indices $(m_1, m_2)$ are uniquely determined by the
$2\times2$ linear system

\begin{equation}
  B(\theta)\,\mathbf{m} = \mathbf{v},
  \label{eq:solve-overview}
\end{equation}

where $B(\theta) = R(\theta)\,G$ is the rotated top-layer basis matrix and
$R(\theta)$ is the 2D rotation matrix for twist angle $\theta$. A
coincidence occurs if and only if $\mathbf{m}$ is close to an integer vector,
i.e.\ both fractional residuals $|m_i - \mathrm{round}(m_i)|$ fall below a
tolerance $\delta$. This reduces the four-index
$\mathcal{O}(N^4)$ brute-force search to a two-index
$\mathcal{O}(N^2)$ enumeration over $(n_1, n_2)$, with one
$2\times2$ linear solve per candidate. The detailed mathematical formulation
and algorithm pseudocode are given in Section~\ref{sec:methods}.

\subsection{Acceptance and validation}

Accepted candidate vectors are further characterized by a Cartesian residual
metric,
\begin{equation}
  \delta_{\mathrm{vec}} = \| B(\theta)\,\hat{m} - A\,n \|_2,
\end{equation}
where $\hat{m} = \mathrm{round}(\mathbf{m}) \in \mathbb{Z}^2$. This
quantity, reported in \AA\ for each match, measures how far the
rounded top-layer vector is from the exact bottom-layer supercell
vector. It serves as a direct quality indicator: a smaller
$\delta_{\mathrm{vec}}$ corresponds to a more nearly perfect
commensurate structure. For production-quality DFT or MD inputs we
recommend the default tolerance $\delta = 10^{-4}$, which gives
$\delta_{\mathrm{vec}} < 0.001$~\AA\ — well below typical DFT force
convergence thresholds.

\subsection{Extension to multilayer systems}

Generalization to three or more layers is performed sequentially. After
finding a commensurate supercell for layers 1 and 2 at angle $\theta_1$,
the identified supercell vectors $(\mathbf{A}_1^{\mathrm{sc}},
\mathbf{A}_2^{\mathrm{sc}})$ serve as the effective reference lattice for
the next search. A new coincidence search is then performed between these
supercell vectors and the third layer primitive lattice rotated by $\theta_2$
relative to the bottom layer, yielding a globally periodic trilayer supercell.
This procedure extends naturally to arbitrarily many layers with independent
twist angles, which is a capability not available in existing bilayer-focused
tools.

\subsection{Atomic structure generation}

Once supercell vectors are identified, the atomic structure is assembled in
three stages. First, each monolayer is replicated over a sufficiently large
supercell to ensure that the target moiré cell is fully contained. Second,
each layer is rotated about the $z$-axis by the appropriate angle and offset
in $z$ by the interlayer spacing. Third, atoms belonging to exactly one
moiré supercell are selected using a fractional-coordinate folding algorithm:
all replicated atom positions are mapped to fractional coordinates with
respect to the supercell basis, folded into $[0,1)^2$, and deduplicated.
This approach is $\mathcal{O}(N)$ in the number of replicated atoms, is
exact for the bottom layer, and handles boundary atoms correctly without any
polygon-clipping tolerance tuning. The resulting structure is written to disk
in VASP POSCAR or LAMMPS atomic-style data format via the Atomic Simulation
Environment (ASE) \cite{Larsen.2017.JPCM}.

\section{Methods}
\label{sec:methods}

We detect commensurate moiré superlattices by rotating one layer and testing coincidence between the two Bravais lattices via a solve–and–round criterion, optionally augmented with a residual check.

\subsection{Lattice setup}
Let $A \in \mathbb{R}^{2\times 2}$ denote the primitive basis of the reference (bottom) layer, and let $G \in \mathbb{R}^{2\times 2}$ be the primitive basis of the second (top) layer prior to rotation. Columns of $A$ and $G$ give the Cartesian primitive vectors.

\subsection{Rotation sweep}
We sample twist angles $\theta$ uniformly in a user-specified interval $[\theta_{\min},\,\theta_{\max})$ with step $\Delta\theta$. For each $\theta$ we construct the rotation
\begin{equation}
R(\theta)=
\begin{bmatrix}
\cos\theta & -\sin\theta\\
\sin\theta & \cos\theta
\end{bmatrix},
\qquad
B(\theta) \;:=\; R(\theta)\,G.
\label{eq:rotation}
\end{equation}

\subsection{Bottom-layer enumeration}
We enumerate integer indices on a square grid,
\begin{equation}
n_1, n_2 \in [n_{\min},\, n_{\max})\cap \mathbb{Z},
\end{equation}
excluding the origin. Each index pair $n=(n_1,n_2)^\top$ maps to a Cartesian lattice vector of the reference layer,
\begin{equation}
\mathbf{v}(n)\;=\; A\,n.
\label{eq:bottom-vector}
\end{equation}

\subsection{Coincidence-site test (solve–and–round)}
For each $\mathbf{v}(n)$ we solve the $2\times 2$ system
\begin{equation}
B(\theta)\,\mathbf{m} \;=\; \mathbf{v}(n)
\label{eq:solve}
\end{equation}
for $\mathbf{m}\in\mathbb{R}^2$ using a direct linear solve. We declare a coincidence-site match when both components of $\mathbf{m}$ lie within a user-tuned tolerance $\delta$ of the nearest integers:
\begin{equation}
\left| m_i - \mathrm{round}(m_i)\right| < \delta \quad (i=1,2).
\label{eq:fractional-tol}
\end{equation}
For accepted matches we record the tuple $\bigl(\theta,\,n,\,\hat m,\,\mathbf{v}(n)\bigr)$, where $\hat m :=\mathrm{round}(\mathbf{m})\in\mathbb{Z}^2$.

\subsection{Residual-based acceptance}
In addition to the fractional-part check \eqref{eq:fractional-tol}, the method can apply a geometric residual test to guard against false positives due to numerical round-off or ill-conditioning of $B(\theta)$.  
Given the rounded integer vector $\hat{m}$, we compute the relative residual
\begin{equation}
\rho = \frac{\|B(\theta)\,\hat{m} - \mathbf{v}(n)\|_2}{\|\mathbf{v}(n)\|_2}.
\label{eq:residual}
\end{equation}
A match is retained only if $\rho < \varepsilon_{\mathrm{res}}$, where $\varepsilon_{\mathrm{res}}$ is a user-defined residual tolerance. This ensures that the rounded integer solution, when mapped back through $B(\theta)$, reconstructs the target vector $\mathbf{v}(n)$ to within the prescribed accuracy.

\subsection{Optional angular filtering}
For a given $\theta$, let $\{\mathbf{v}_j\}$ be the set of accepted moiré vectors from \eqref{eq:bottom-vector}. To prefer approximately symmetric supercells, we optionally retain only pairs $(\mathbf{v}_i,\mathbf{v}_j)$ whose mutual angle
\begin{equation}
\phi_{ij} \;=\; \cos^{-1}\!\left(\frac{\mathbf{v}_i^\top \mathbf{v}_j}{\|\mathbf{v}_i\|\,\|\mathbf{v}_j\|}\right)
\end{equation}
lies within a tolerance of a target angle $\phi_{\mathrm{target}}$ (e.g., $\sim 60^\circ$ for near-hexagonal moiré cells).

\subsection{Outputs}
For each sampled angle, the procedure returns (i) the accepted bottom-layer vectors $\mathbf{v}_j$, (ii) their indices $n=(n_1,n_2)$ in the bottom lattice and the rounded indices $\hat m=(m_1,m_2)$ in the rotated top lattice, and (iii) optionally, the subset of vector pairs that satisfy the angular filter. The residual criterion \eqref{eq:residual} is optional and can be applied downstream after the initial match detection.

\DontPrintSemicolon
\SetKwInOut{KwIn}{Input}
\SetKwInOut{KwOut}{Output}
\SetKw{KwTo}{to}
\SetKw{KwBy}{by}
\SetKw{Continue}{continue}

\begin{algorithm}[H]
\caption{Moiré search via solve--and--round }
\KwIn{Bases $A,G\in\mathbb{R}^{2\times 2}$; angle sweep $[\theta_{\min},\theta_{\max})$ with step $\Delta\theta$; index window $[n_{\min},n_{\max})$; fractional tolerance $\delta$; residual tolerance $\varepsilon_{\mathrm{res}}$; flag \texttt{residual\_enabled}.}
\KwOut{For each $\theta$, accepted matches $\mathcal{M}_\theta$ as tuples $(\mathbf{v},\,n,\,\hat m)$.}

\For{$\theta \leftarrow \theta_{\min}$ \KwTo $\theta_{\max}$ \KwBy $\Delta\theta$}{
  $B \leftarrow R(\theta)\,G$\;
  $\mathcal{M}_\theta \leftarrow \varnothing$\;
  \For{$n_1 \leftarrow n_{\min}$ \KwTo $n_{\max}-1$}{
    \For{$n_2 \leftarrow n_{\min}$ \KwTo $n_{\max}-1$}{
      \If{$(n_1,n_2)=(0,0)$}{\Continue}
      $\mathbf{v} \leftarrow A\,[n_1,n_2]^\top$\;
      Solve $B\,\mathbf{m}=\mathbf{v}$ for $\mathbf{m}\in\mathbb{R}^{2}$\;
      \If{$|m_1-\mathrm{round}(m_1)|<\delta$ \textbf{and} $|m_2-\mathrm{round}(m_2)|<\delta$}{
        $\hat m \leftarrow \mathrm{round}(\mathbf{m})$\;
        \eIf{\texttt{residual\_enabled}}{
          $\rho \leftarrow \dfrac{\|B\,\hat m - \mathbf{v}\|_2}{\max(\|\mathbf{v}\|_2,\,10^{-16})}$\;
          \If{$\rho < \varepsilon_{\mathrm{res}}$}{
            $\mathcal{M}_\theta \leftarrow \mathcal{M}_\theta \cup \{(\mathbf{v},(n_1,n_2),\hat m)\}$\;
          }
        }{
          $\mathcal{M}_\theta \leftarrow \mathcal{M}_\theta \cup \{(\mathbf{v},(n_1,n_2),\hat m)\}$\;
        }
      }
    }
  }
  \tcp{Optional: filter $\mathcal{M}_\theta$ by inter-vector angle near a target $\phi_{\mathrm{target}}$}
  \Return $(\theta,\mathcal{M}_\theta)$
}
\end{algorithm}

\section{Software Design and Implementation}
\label{sec:software}

\subsection{Package Architecture}

MLM is implemented in Python~3 and organized as an installable package under
\texttt{src/MLM/}. The package comprises three principal modules:
\texttt{match}, \texttt{moire\_lattice\_vector\_finder}, and
\texttt{structure\_writer}, each encapsulating a distinct stage of the
workflow: lattice-match discovery, high-throughput index search, and atomic
structure generation, respectively. Supplementary Jupyter notebooks in
\texttt{notebooks/} provide interactive interfaces for parameter exploration
and result visualization. Pre-computed candidate structures for representative
material systems (graphene, MoS$_2$, SrTiO$_3$, and PTO/BTO heterostructures)
are stored as serialized DataFrame objects under \texttt{moire\_structures/}.
The full dependency list is given in Table~\ref{tab:deps}.

\begin{table}[h]
\centering
\caption{Python dependencies of the MLM package.}
\label{tab:deps}
\begin{tabular}{ll}
\toprule
Package & Purpose \\
\midrule
NumPy   & Linear algebra and vectorised array operations \\
Polars  & High-performance DataFrame storage of match results \\
Numba   & JIT compilation of compute-intensive kernels \\
ASE     & Atomic structure I/O (VASP, LAMMPS formats) \\
SciPy   & KDTree for boundary atom deduplication \\
Pandas  & Intermediate DataFrame handling in structure writer \\
Matplotlib & Optional visualisation \\
\bottomrule
\end{tabular}
\end{table}

\subsection{Mathematical Framework: Change-of-Basis Formulation}
\label{sec:cob}

The matching problem is formulated within a rigorous change-of-basis algebra
operating across four coordinate systems:
\begin{itemize}
  \item \textbf{Iota (I)}: the standard Cartesian frame.
  \item \textbf{Alpha (A)}: the lattice frame of the bottom layer, with
    change-of-basis matrix $A$ whose columns are the primitive vectors
    $\mathbf{a}_1$, $\mathbf{a}_2$.
  \item \textbf{Gamma (G)}: the lattice frame of the top layer, with
    change-of-basis matrix $G$.
  \item \textbf{Beta (B)}: the rotated top-layer frame,
    $B = R(\theta)G$, where $R(\theta)$ is the 2D rotation matrix for
    twist angle $\theta$.
\end{itemize}

A lattice vector has integer components only when expressed in its own frame.
A point in the bottom layer, with integer indices $(n_1, n_2)$ in the Alpha
frame, is also a lattice point of the rotated top layer if and only if its
representation in the Beta frame,
\begin{equation}
  \mathbf{p}_{\beta} = B^{-1} A \begin{pmatrix} n_1 \\ n_2 \end{pmatrix},
\end{equation}
has integer-valued components. The residuals of $\mathbf{p}_\beta$ from the
nearest integer vector measure the lattice mismatch; a match is accepted when
both residuals fall below the user-supplied tolerance $\delta_{\mathrm{tol}}$.

\subsection{Lattice-Match Search (\texttt{match} module)}

The primary search is implemented in \texttt{scan(A, G, $\theta$, tol,
xlim, ylim)}. For a given rotation angle $\theta$, the function:
\begin{enumerate}
  \item Constructs the rotated basis matrix $B = R(\theta)G$.
  \item Enumerates all bottom-layer lattice points $\{(n_1,n_2)\}$ with
    indices in the range \texttt{[nmin, nmax]}, excluding the origin.
  \item Transforms each point to the Beta frame via
    $\mathbf{p}_\beta = B^{-1}A\,\mathbf{n}$ using a batched
    \texttt{numpy.linalg.solve} call.
  \item Identifies commensurate points whose fractional residuals satisfy
    $|f_i| < \delta_{\mathrm{tol}}$.
  \item Records the top-layer indices $(m_1, m_2)$ (rounded Beta coordinates),
    the mismatch vector magnitude $\delta_{\mathrm{vec}} =
    \|B\hat{m} - A n\|_2$, and the match-point position in the Iota frame.
  \item Returns up to 25 candidate matches sorted by lattice vector norm.
\end{enumerate}

The outer driver \texttt{run()} iterates over a discrete angle grid
$[\theta_{\min},\theta_{\max})$ with step $\Delta\theta$, aggregating
results into a Polars DataFrame for efficient downstream filtering.

Post-processing is handled by \texttt{filtermatches()}, which, for each
twist angle, computes pairwise dot products between candidate match vectors
and selects pairs whose mutual angle deviates from a prescribed target
(e.g.\ $60^\circ$, $90^\circ$, or $120^\circ$) by less than a relative
tolerance. This step identifies physically meaningful unit-cell geometries —
rectangular, rhombic, or hexagonal moiré supercells — and is the key step
for selecting valid two-vector supercell bases.

\subsection{High-Throughput Parallelized Search
  (\texttt{moire\_lattice\_vector\_finder} module)}

For large-scale surveys over dense angle grids, a second search engine is
provided in \texttt{find\_index()}. The innermost kernel,
\texttt{find\_index\_chunk()}, performs a brute-force four-index loop over
$(n_1, n_2, m_1, m_2)$:
\begin{equation}
  \mathbf{v}_1 = n_1\mathbf{a}_1 + n_2\mathbf{a}_2, \qquad
  \mathbf{v}_2 = m_1 R(\theta)\mathbf{b}_1 + m_2 R(\theta)\mathbf{b}_2,
\end{equation}
accepting pairs for which $|\mathbf{v}_1 - \mathbf{v}_2| < \delta_{\mathrm{tol}}$.
The kernel is compiled with Numba's \texttt{@njit} decorator for
near-native execution speed. Parallelisation over the angle grid is achieved
via \texttt{multiprocessing.Pool}, distributing contiguous angle chunks across
all available CPU cores. Matched index pairs are further filtered by the angle
between the two superlattice vectors to enforce a target unit-cell geometry.

\subsection{Atomic Structure Generation (\texttt{structure\_writer} module)}

Once commensurate lattice vectors are identified, the corresponding atomic
structure is assembled through a three-stage pipeline.

\textbf{Stage 1 — Input parsing.}
VASP POSCAR files are read by \texttt{read\_vasp()}, extracting the
$3\times3$ lattice tensor, elemental species identifiers, and fractional or
Cartesian atomic positions.

\textbf{Stage 2 — Supercell replication.}
\texttt{replicate\_atoms()} tiles the unit cell over a range
$[-N_a, N_a) \times [-N_b, N_b) \times [0, N_c)$ of lattice translations.
Displacement vectors for all images are computed with a vectorised NumPy
outer sum, avoiding explicit Python loops over atoms.

\textbf{Stage 3 — Moiré unit-cell extraction.}
Atoms belonging to exactly one moiré supercell period are selected by
\texttt{select\_atoms\_fractional()}, which maps each atom's Cartesian
$(x,y)$ position to fractional coordinates with respect to the supercell
vectors $\mathbf{A}_1^{\mathrm{sc}}, \mathbf{A}_2^{\mathrm{sc}}$ via a
$2\times2$ linear solve:
\begin{equation}
  \begin{pmatrix} f_1 \\ f_2 \end{pmatrix}
  = \bigl[\mathbf{A}_1^{\mathrm{sc}} \mid \mathbf{A}_2^{\mathrm{sc}}\bigr]^{-1}
    \begin{pmatrix} x \\ y \end{pmatrix}.
\end{equation}
Fractional coordinates are folded into $[0,1)^2$ by $f \to f \bmod 1$, and
periodic images are deduplicated by a two-pass procedure. Pass 1 uses integer
binning with bin width $\sim\ell_{\mathrm{bond}}/(3|\mathbf{A}_1^{\mathrm{sc}}|)$
to collapse $>99\%$ of duplicate images in a single vectorised operation.
Pass 2 uses a periodic-distance KD-tree search (scipy \texttt{KDTree} with
\texttt{boxsize=1}) to catch the small fraction of duplicate pairs that
straddle bin boundaries. The combined procedure is $\mathcal{O}(N_{\mathrm{atom}})$
and requires no manual polygon-clipping tolerances, making it robust at all
twist angles from very small ($\sim 0.08^\circ$) to large ($>45^\circ$).

The bin width is set adaptively based on the estimated primitive bond length
(determined from the 5th percentile of nearest-neighbour distances in the
replicated structure) and the per-structure mismatch residual
$\delta_{\mathrm{vec}}$. This ensures that periodic images — which differ by
at most $\delta_{\mathrm{vec}}/|\mathbf{A}_1^{\mathrm{sc}}|$ in fractional
coordinates — are correctly merged, while distinct crystallographic sites
remain resolved.

Structures for each layer are assembled independently, merged with an
explicit interlayer separation, and written to disk via ASE
\cite{Larsen.2017.JPCM}: \texttt{write\_vasp\_ase()} produces VASP~5
POSCAR files with atoms sorted by species, and \texttt{write\_lammps\_ase()}
produces LAMMPS data files in the atomic style with explicit mass sections,
suitable for use with LAMMPS \cite{Thompson.2022.CPC}.

\section{Results and Examples}
\label{sec:results}

We demonstrate MLM on five representative material systems: bilayer graphene,
bilayer MoS$_2$, trilayer MoS$_2$, bilayer SrTiO$_3$, and a
PbTiO$_3$/SrTiO$_3$ oxide heterostructure. For each system we report the
commensurate twist angles found within a prescribed range, the corresponding
supercell vectors , and the mismatch residual
$\delta_{\mathrm{vec}}$. All structures were generated using the default
tolerance $\delta = 5 \times 10^{-4}$ unless stated otherwise.

\subsection{Validation: Bilayer Graphene}
\label{sec:graphene}

Commensurate twist angles for bilayer graphene are well-established in the
literature \cite{Feuerbacher.2021.ACSAFaA,Cao.2018.N,Cao.2018.Na} and
serve as a natural validation target. Graphene has a hexagonal primitive cell
with lattice constant $a = 2.46$~\AA, with two atoms per unit cell. We ran
MLM with a search range $n_{\min} = -10$, $n_{\max} = 10$ over angles
$0.1^\circ$ to $30.0^\circ$ in steps of $0.01^\circ$.

Table~\ref{tab:graphene} lists a representative set of commensurate
configurations found by MLM. The angles are in excellent agreement with
analytically derived values from coincidence-lattice theory
\cite{Feuerbacher.2021.ACSAFaA}, confirming the correctness of the algorithm.

\begin{table}[h]
\centering
\caption{Selected commensurate supercells for bilayer graphene ($a = 2.46$~\AA,
  $\delta = 5 \times 10^{-4}$). Columns: twist angle $\theta$,  supercell vector norm
  $|\mathbf{A}_1^{\mathrm{sc}}|$, and
  Cartesian mismatch $\delta_{\mathrm{vec}}$.}
\label{tab:graphene}
\begin{tabular}{ccccccr}
\toprule
$\theta$ (deg) &
  $|\mathbf{A}_1^{\mathrm{sc}}|$ (\AA) &
  $\delta_{\mathrm{vec}}$ (\AA) \\
\midrule
21.79  &  6.5278  & $ 3.6 \times 10^{-4}$ \\
13.17  &  10.7547 & $ 6.7 \times 10^{-4}$ \\
9.43   &  15.0079 & $ 2.0 \times 10^{-6}$ \\
16.43  &  17.2710 & $ 1.0 \times 10^{-3}$ \\
7.34   &  19.2701 & $ 3.3 \times 10^{-4}$ \\
\bottomrule
\end{tabular}
\end{table}

\begin{figure}[H]
\centering
\includegraphics[width=0.55\linewidth]{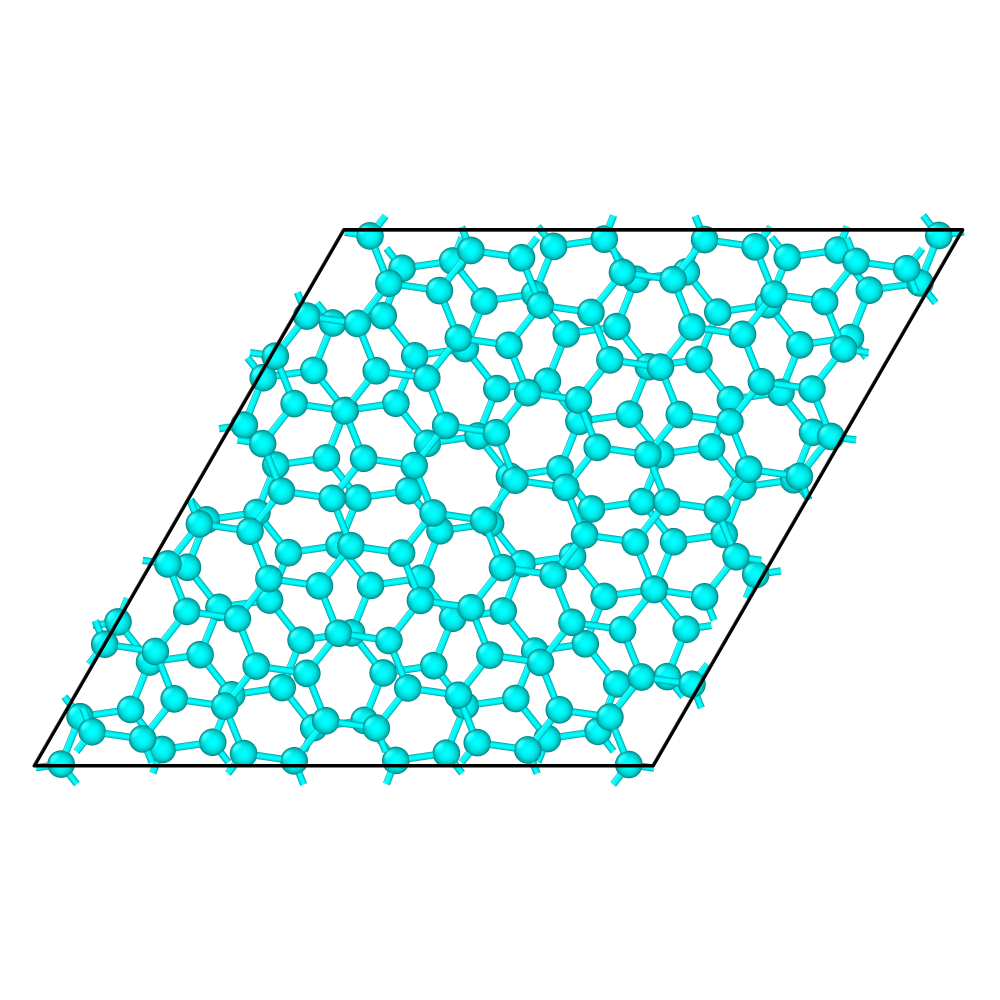}
\caption{Top view of the $16.43^\circ$ commensurate bilayer graphene supercell
  . Carbon atoms from the bottom and top layer are shown in cyan. The parallelogram outlines the moiré unit cell defined
  by the two supercell vectors $\mathbf{A}_1^{\mathrm{sc}}$ and
  $\mathbf{A}_2^{\mathrm{sc}}$.}
\label{fig:graphene}
\end{figure}

Figure~\ref{fig:graphene} shows the moiré pattern for the $16.43^\circ$
configuration, illustrating the alternating AA and AB stacking regions
characteristic of the hexagonal moiré superlattice.

\subsection{Bilayer MoS\texorpdfstring{$_2$}{2}}
\label{sec:mos2}

MoS$_2$ is a transition metal dichalcogenide (TMDC) with a hexagonal lattice
constant $a = 3.17$~\AA\ and three atoms per primitive cell (one Mo, two S).
Twisted bilayer MoS$_2$ has been studied extensively both experimentally and
computationally \cite{Arnold.2023.2M,Naik.2019.TJoPCC,Zande.2014.NL}. We
searched for commensurate cells over $1.0^\circ$ to $30^\circ$ with a search
range of $n_{\min} = -10$, $n_{\max} = 10$

Table~\ref{tab:mos2} lists selected commensurate configurations. Because
MoS$_2$.

\begin{table}[h]
\centering
\caption{Selected commensurate supercells for bilayer MoS$_2$ ($a = 3.18$~\AA,
  $\delta = 5 \times 10^{-4}$).}
\label{tab:mos2}
\begin{tabular}{cccccc}
\toprule
$\theta$ (deg) &
  $|\mathbf{A}_1^{\mathrm{sc}}|$ (\AA) &
  $\delta_{\mathrm{vec}}$ (\AA) \\
\midrule
21.79  &  8.3764  & $ 4.7 \times 10^{-4}$ \\
27.80  &  11.4151 & $ 8.4 \times 10^{-4}$ \\
17.89  &  17.6275 & $ 1.0 \times 10^{-3}$ \\

\bottomrule
\end{tabular}
\end{table}

\begin{figure}[H]
\centering
\includegraphics[width=0.55\linewidth]{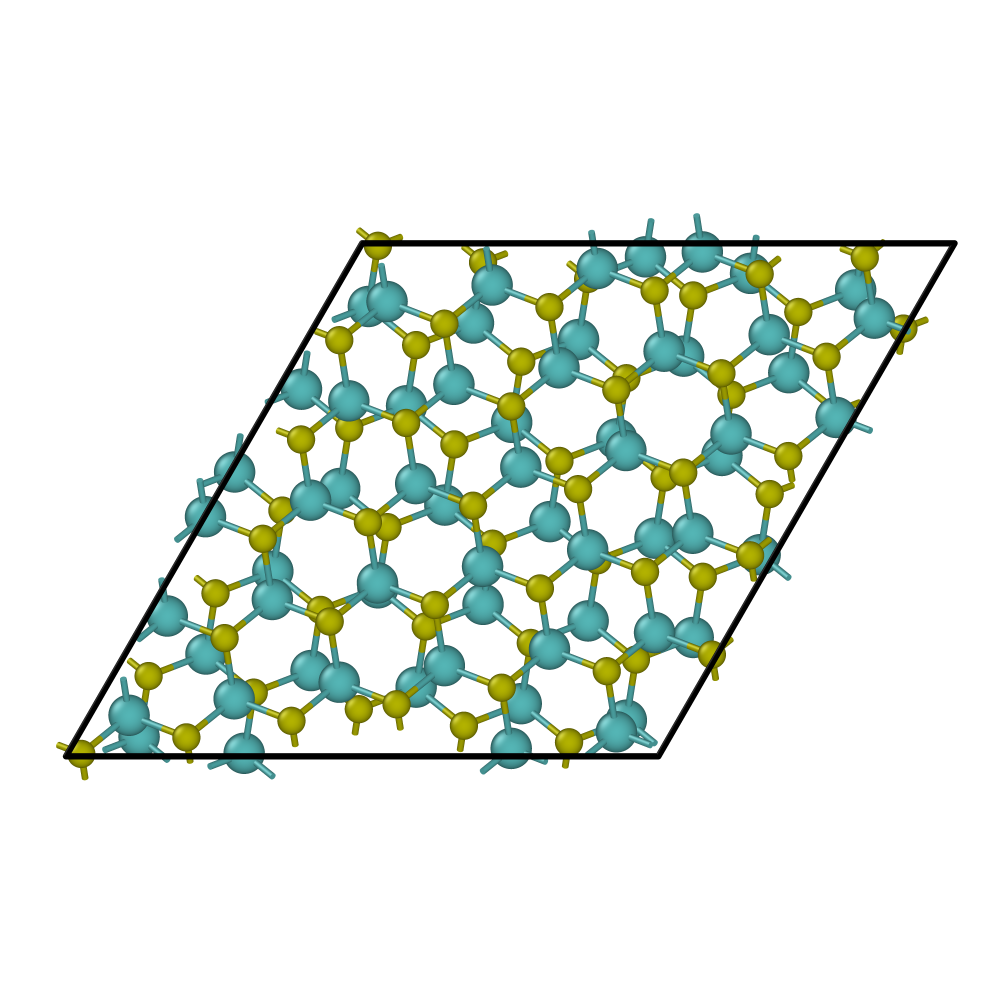}
\caption{Top view of the $17.89^\circ$ commensurate bilayer MoS$_2$ supercell
  . Mo atoms are shown in teal, S atoms in yellow.}
\label{fig:mos2}
\end{figure}

\subsection{Trilayer MoS\texorpdfstring{$_2$}{2}}
\label{sec:trilayer}

The extension to three layers is a key capability of MLM not available in
existing bilayer-focused codes. A trilayer stack is constructed by first
identifying a commensurate bilayer supercell for layers~1 and~2 at angle
$\theta_1$, and then using the resulting supercell vectors as the reference
for a second coincidence search between the supercell and layer~3 at angle
$\theta_2$.

\begin{figure}[H]
\centering
\includegraphics[width=0.65\linewidth]{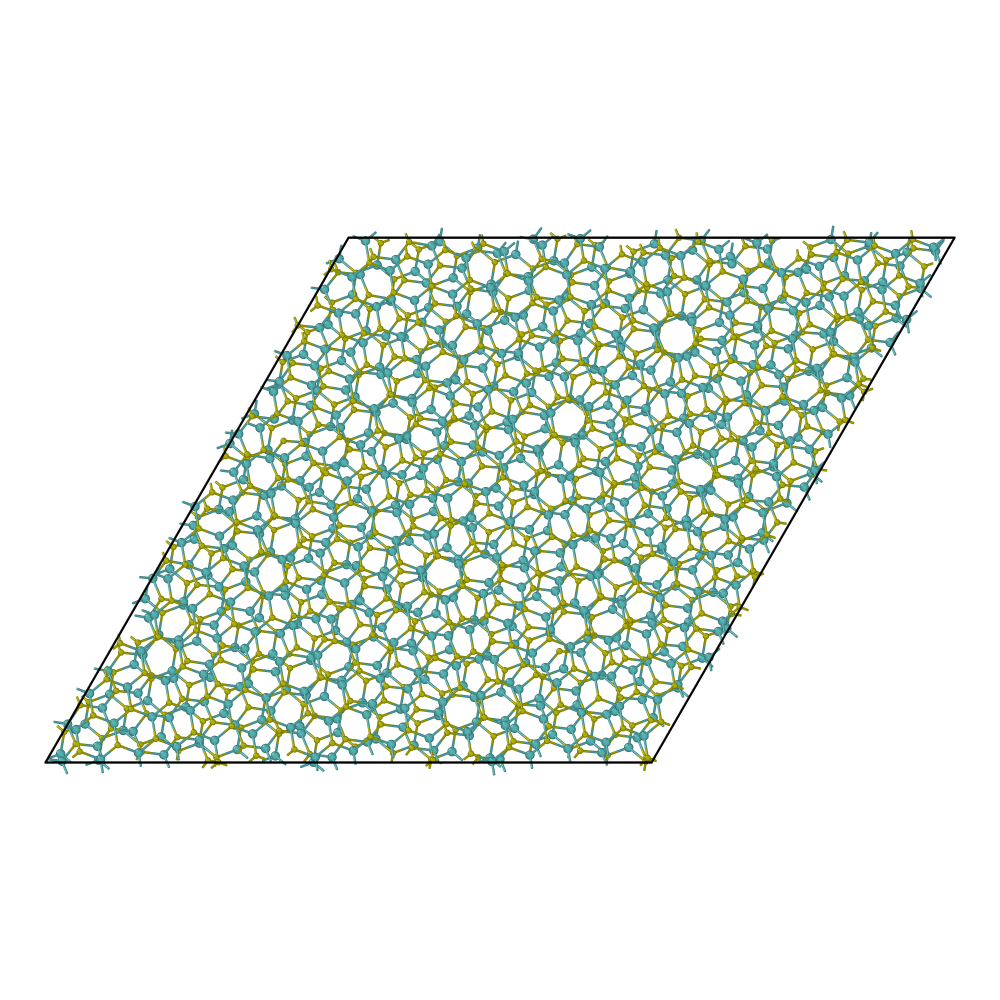}
\caption{Top view of a trilayer MoS$_2$ commensurate supercell with
  $\theta_1 = 27.8^\circ$ between layers 1 and 2, and
  $\theta_2 = 40.97^\circ$ between layers 1 and 3. }
\label{fig:trilayer}
\end{figure}

Figure~\ref{fig:trilayer} shows the moiré pattern for a trilayer MoS$_2$
configuration with $\theta_1 = 27.8^\circ$ and $\theta_2 = 40.97^\circ$.
The three-layer stack produces a more complex spatial modulation of the
local stacking environment, which can give rise to hierarchical moiré
patterns and novel flat-band features \cite{frontiers_moire_2022,Du.2023.S}.

\subsection{Bilayer SrTiO\texorpdfstring{$_3$}{3}}
\label{sec:sto}

SrTiO$_3$ (STO) is a perovskite oxide with ,
$a = b = 3.905$~\AA, and five atoms per primitive cell. Unlike graphene or
MoS$_2$, the STO lattice is not hexagonal, which prevents the use of
symmetry-specific angle formulas. MLM handles this directly because the
solve-and-round algorithm makes no assumptions about lattice symmetry.

We searched for commensurate bilayer STO cells over $1^\circ$ to $90^\circ$.
The results demonstrate that MLM correctly identifies commensurate angles for
square-symmetry lattices  
$\approx 36.87^\circ$ (arctan$(3/4)$) configurations that are well-known in
the perovskite thin-film literature \cite{SanchezSantolino2024,Lee2024}.

\begin{figure}[H]
\centering
\includegraphics[width=0.55\linewidth]{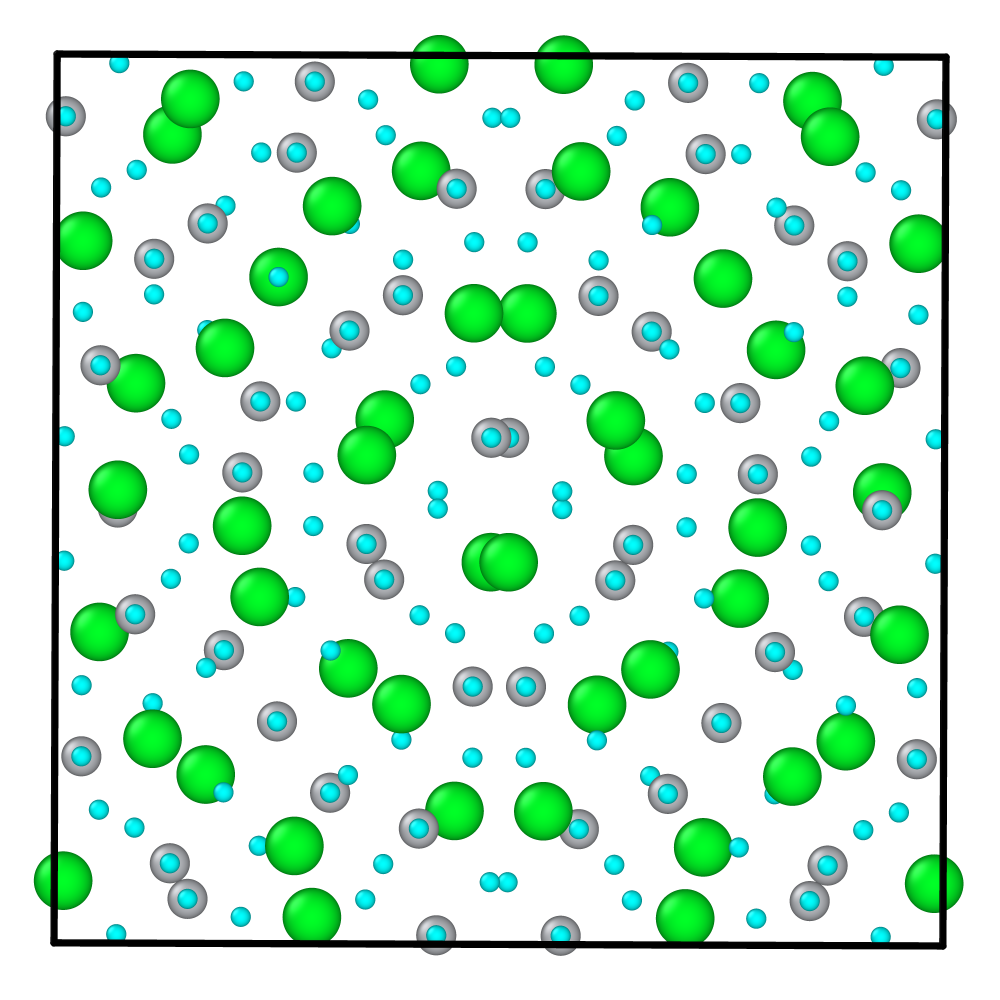}
\caption{Top view of a commensurate bilayer SrTiO$_3$ supercell at a
  representative twist angle of $ 16.26^\circ$. Sr, Ti, and O atoms are shown in green, blue,
  and cyan respectively.}
\label{fig:sto}
\end{figure}

\subsection{PbTiO\texorpdfstring{$_3$}{3}/SrTiO\texorpdfstring{$_3$}{3}
  Heterostructure}
\label{sec:pto-bto}

The PbTiO$_3$ (PTO)/SrTiO$_3$ (STO) system is a prototypical ferroelectric
oxide heterostructure. PTO and STO both adopt the perovskite structure but
have different lattice constants: $a_{\mathrm{PTO}} = 3.880$~\AA\ and
$a_{\mathrm{STO}} = 3.91$~\AA\ (a $\sim 0.7\%$ mismatch). Because the two
lattice constants are incommensurate. MLM finds the best
\textit{approximately} commensurate cells, quantified by
$\delta_{\mathrm{vec}}$.

We generated commensurate PTO/STO heterostructure cells over the full
$1^\circ$ to $90^\circ$ range. We found few possible structures at the twist of $ 7.13^\circ$, $ 17.74^\circ$, $ 28.99^\circ$, $ 29.74^\circ$, $ 35.02^\circ$, and $ 40.36^\circ$

\begin{figure}[H]
\centering
\includegraphics[width=0.60\linewidth]{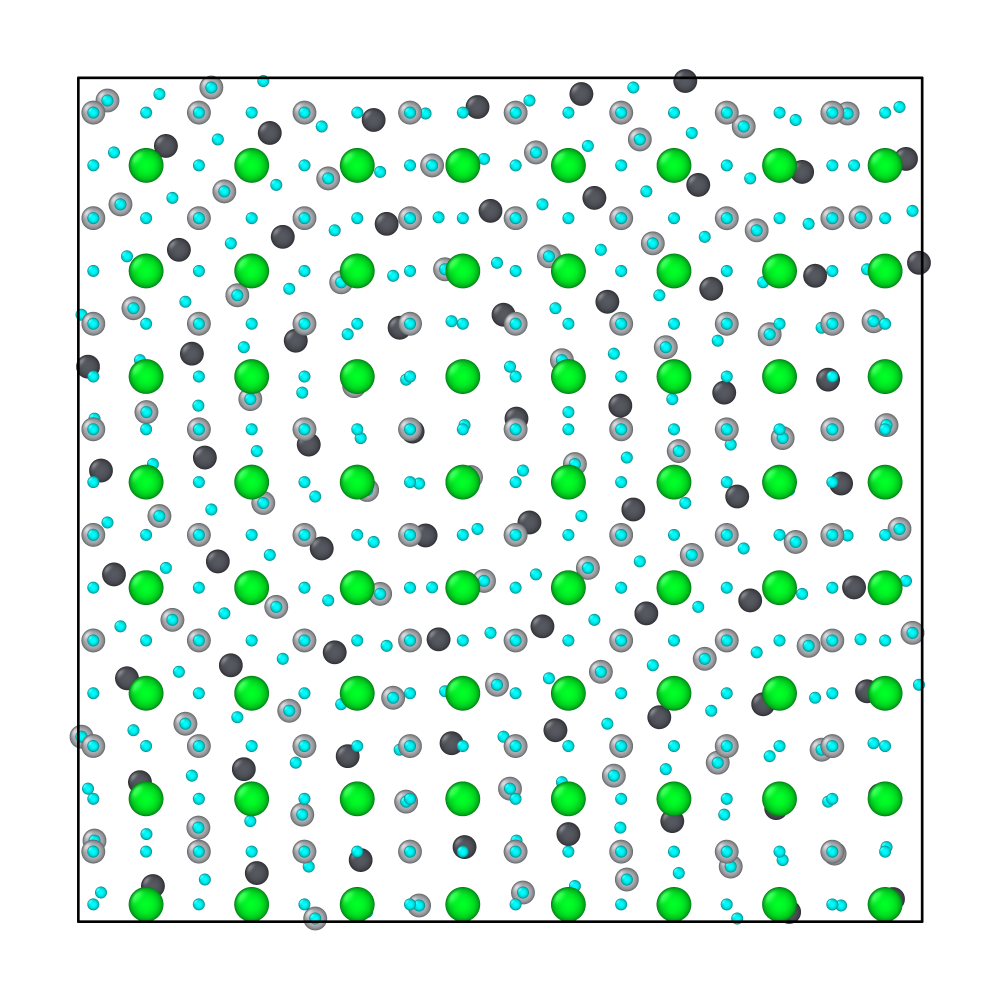}
\caption{Top view of an approximately commensurate PbTiO$_3$/SrTiO$_3$
  heterostructure supercell at twist of $ 7.13^\circ$}
\label{fig:pto-bto}
\end{figure}

\section{Performance Comparison}
\label{sec:performance}

\subsection{Algorithmic complexity}

The solve-and-round strategy employed by MLM achieves a qualitative reduction
in computational complexity compared to brute-force coincidence search. For a
search range of $\pm N$ primitive lattice vectors:

\begin{itemize}
  \item \textbf{Brute-force} (as used, for example, in TWISTER
    \cite{Naik.2022.CPC}): all four-index combinations $(n_1, n_2, m_1, m_2)$
    must be tested, giving $(2N+1)^4$ candidate evaluations per twist angle.
    This grows as $\mathcal{O}(N^4)$.

  \item \textbf{MLM solve-and-round}: only the $(2N+1)^2 - 1$ bottom-layer
    vectors $(n_1, n_2)$ are enumerated; for each, the top-layer indices
    $(m_1, m_2)$ are computed by a single $2\times2$ linear solve. This
    grows as $\mathcal{O}(N^2)$.
\end{itemize}

The two-order-of-magnitude reduction in the exponent is independent of the
material system or lattice type, and is achieved simply by computing the
top-layer indices instead of guessing them.

\subsection{Benchmark}

To quantify the practical speedup, we timed both algorithms on the same
problem: finding all commensurate bilayer graphene supercells over
$1^\circ$–$30^\circ$ in steps of $0.1^\circ$, on a single CPU core. The
brute-force implementation was fully vectorised with NumPy (the same strategy
used in TWISTER) to ensure a fair comparison. Both algorithms found
the same set of commensurate structures.

\begin{figure}[H]
\centering
\includegraphics[width=0.60\linewidth]{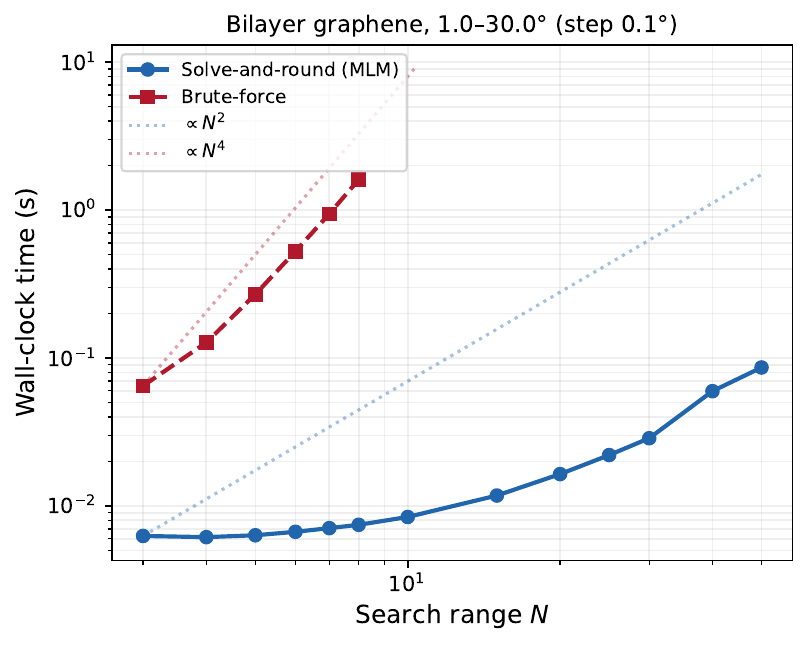}
\caption{Wall-clock time vs.\ search range $N$ for the solve-and-round
  (MLM, blue circles) and brute-force (red squares) algorithms on
  bilayer graphene, $1^\circ$–$30^\circ$, step $0.1^\circ$, single CPU core.
  Dashed reference lines show $N^2$ and $N^4$ slopes.}
\label{fig:benchmark}
\end{figure}

Figure~\ref{fig:benchmark} shows wall-clock time as a function of search
range $N$ on a log-log scale. The fitted scaling exponents from linear
regression in log-log space are 0.93 for MLM and 3.32 for brute-force
(Table~\ref{tab:benchmark}). These are somewhat below the theoretical
values of 2 and 4 because at these small $N$ values the fixed overhead of
iterating over 290 angles dominates; the $N$-dependent kernel cost grows
relative to this overhead as $N$ increases, which is reflected in the
speedup growing monotonically from $10\times$ at $N=3$ to $216\times$ at
$N=8$.

\begin{table}[h]
\centering
\caption{Benchmark timing summary for bilayer graphene
  ($1^\circ$–$30^\circ$, step $0.1^\circ$), single CPU core.
  Columns: search range $N$, number of bottom-layer candidates per angle
  for MLM ($\approx(2N+1)^2$), wall-clock time for MLM and brute-force,
  and speedup ratio.}
\label{tab:benchmark}
\begin{tabular}{rrrrrr}
\toprule
$N$ & MLM candidates & MLM (s) & Brute-force (s) & Speedup \\
\midrule
3  &   48  & 0.006 & 0.065  &  10$\times$ \\
4  &   80  & 0.006 & 0.127  &  21$\times$ \\
5  &  120  & 0.006 & 0.270  &  43$\times$ \\
6  &  168  & 0.007 & 0.527  &  79$\times$ \\
7  &  224  & 0.007 & 0.943  & 133$\times$ \\
8  &  288  & 0.008 & 1.610  & 216$\times$ \\
\midrule
\multicolumn{2}{l}{Fitted exponent} & 0.93 & 3.32 & \\
\bottomrule
\end{tabular}
\end{table}

\subsection{Practical implications}

The $\mathcal{O}(N^2)$ scaling has two practical consequences:

\begin{enumerate}
  \item \textbf{Larger search ranges at the same cost.} At a given computation
    time budget, MLM can search supercell sizes up to $\sim N^2$ times larger
    than brute-force. This is critical for small twist angles ($<1^\circ$)
    where supercell vectors can exceed $100\times$ the primitive lattice
    constant, requiring $N > 100$.

  \item \textbf{Single-core efficiency.} Because the $\mathcal{O}(N^2)$
    kernel is fast enough on a single core, MLM does not require MPI or GPU
    acceleration for typical angle ranges. For very dense scans, the outer
    loop over twist angles can be trivially parallelised using
    \texttt{multiprocessing} (Section~\ref{sec:software}).
\end{enumerate}

It should be noted that TWISTER \cite{Naik.2022.CPC} offers MPI
parallelisation of its brute-force loop, which provides a practical speedup
proportional to the number of MPI ranks. This does not change the
$\mathcal{O}(N^4)$ asymptotic scaling, and the MLM single-core performance
remains superior at all $N$ values tested here. Both approaches find
the same commensurate structures; the difference is purely in computational
cost.
\section{Conclusions}
\label{sec:conclusions}

We have presented MLM (Multi-Layer Moir\'{e}), an open-source Python
package for constructing commensurate moiré supercells for twisted multilayer
two-dimensional materials. The main contributions of this work are:

\begin{enumerate}

  \item \textbf{Efficient solve-and-round algorithm.}
    The coincidence-site search is reduced from an $\mathcal{O}(N^4)$
    brute-force enumeration to an $\mathcal{O}(N^2)$ procedure by noting
    that, for each candidate bottom-layer vector $(n_1, n_2)$, the
    corresponding top-layer fractional indices $(m_1, m_2)$ are uniquely
    determined by a $2\times2$ linear solve. Benchmarks on bilayer graphene
    confirm the predicted scaling exponents and demonstrate speedups of
    $10^2$–$10^3\times$ at practical search ranges, enabling the exploration
    of small-angle moiré structures that would be computationally prohibitive
    with brute-force enumeration.

  \item \textbf{Multilayer generalization.}
    The sequential extension to three or more layers is a capability not
    available in existing bilayer-focused codes. Any number of layers with
    independent twist angles can be accommodated, with each additional layer
    requiring only one additional $\mathcal{O}(N^2)$ search. We demonstrated
    this for trilayer MoS$_2$.

  \item \textbf{Lattice-type independence.}
    The algorithm makes no assumptions about hexagonal or identical lattice
    symmetry. We demonstrated correct results for hexagonal (graphene,
    MoS$_2$), square (SrTiO$_3$), and heterostructure (PbTiO$_3$/SrTiO$_3$)
    systems. For heterobilayers with inherent lattice mismatch, the
    $\delta_{\mathrm{vec}}$ metric quantifies how far each found structure is
    from exact commensurability, allowing users to choose an acceptable
    mismatch tolerance for their application.

\end{enumerate}

\subsection*{Future directions}

Several extensions of the present work are natural:

\begin{itemize}

  \item \textit{Per-layer atom selection.} The present implementation selects
    atoms jointly for all layers. Exploiting the bottom/top layer asymmetry
    — where the bottom layer is always exactly periodic and the top layer
    carries the mismatch — would allow per-layer selection with tighter
    deduplication tolerances.

  \item \textit{Direct ASE and pymatgen integration.} A workflow interface
    that accepts ASE \texttt{Atoms} objects or pymatgen \texttt{Structure}
    objects directly, and returns simulation-ready structures in the same
    format, would facilitate integration with existing computational
    materials workflows.

  \item \textit{GUI interface.} An interactive graphical interface for
    parameter exploration and real-time visualization of the moiré pattern
    would lower the barrier for users unfamiliar with Python scripting.
\end{itemize}

\section*{Acknowledgements}
The authors used Claude (Anthropic) to assist with manuscript drafting, editing, and generation of the benchmark comparison script and the fractional coordinate atom selection implementation.

\section*{Data availability}

All example structure files and the scripts used to generate the results
in this paper are available in the MLM repository at
\url{https://github.com/anikeya9/Multi-Layer-Moire-MLM-/tree/master}.

\bibliographystyle{elsarticle-num}
\bibliography{ref}

\end{document}